\documentclass[11pt,a4paper]{article}
\usepackage{amsmath, amsfonts, amssymb}
\usepackage{color} 
\usepackage{authblk} 
\usepackage{hyperref}
\usepackage{graphicx}
\usepackage{wasysym}
\usepackage{epsfig}
\usepackage{multicol}

\newtheorem{theorem}{Theorem}[section]

\newtheorem{proposition}[theorem]{Proposition}
\newtheorem{corollary}[theorem]{Corollary}
\newtheorem{lemma}[theorem]{Lemma}

\newtheorem{remark}[theorem]{Remark}
\newtheorem{conjecture}[theorem]{Conjecture}
\numberwithin{equation}{section}
\def\qed{\hfill $\Box$\medskip}
\def\IR{{\mathbb R}}
\def\IC{{\mathbb C}}
\def\bx{{\bf x}}
\def\by{{\bf y}}
\def\bv{{\bf v}}
\def\bu{{\bf u}}
\def\diag{{\rm diag}\,}

\def\Span{{\rm Span}\,}

\def\tr{{\rm tr}}
\def\bA{{\bf A}}
\newcommand{\mf}[1]{\mbox{\footnotesize{#1}}}

\makeatletter
\newif\if@borderstar
\def\bordermatrix{\@ifnextchar*{%
\@borderstartrue\@bordermatrix@i}{\@borderstarfalse\@bordermatrix@i*}%
}
\def\@bordermatrix@i*{\@ifnextchar[{\@bordermatrix@ii}{\@bordermatrix@ii[()]}}
\def\@bordermatrix@ii[#1]#2{%
\begingroup
\m@th\@tempdima8.75\p@\setbox\z@\vbox{%
\def\cr{\crcr\noalign{\kern 2\p@\global\let\cr\endline }}%
\ialign {$##$\hfil\kern 2\p@\kern\@tempdima & \thinspace %
\hfil $##$\hfil && \quad\hfil $##$\hfil\crcr\omit\strut %
\hfil\crcr\noalign{\kern -\baselineskip}#2\crcr\omit %
\strut\cr}}%
\setbox\tw@\vbox{\unvcopy\z@\global\setbox\@ne\lastbox}%
\setbox\tw@\hbox{\unhbox\@ne\unskip\global\setbox\@ne\lastbox}%
\setbox\tw@\hbox{%
$\kern\wd\@ne\kern -\@tempdima\left\@firstoftwo#1%
\if@borderstar\kern2pt\else\kern -\wd\@ne\fi%
\global\setbox\@ne\vbox{\box\@ne\if@borderstar\else\kern 2\p@\fi}%
\vcenter{\if@borderstar\else\kern -\ht\@ne\fi%
\unvbox\z@\kern-\if@borderstar2\fi\baselineskip}%
\if@borderstar\kern-2\@tempdima\kern2\p@\else\,\fi\right\@secondoftwo#1 $%
}\null \;\vbox{\kern\ht\@ne\box\tw@}%
\endgroup
}
\makeatother

\title{A Note on Parallel  Distinguishability of Two Quantum Operations }

\author[1]{Chi-Kwong Li\thanks{ckli@math.wm.edu }}
\author[2]{Yue Liu\thanks{liuyue09@gmail.com }}
\author[3]{Chao Ma\thanks{machao0923@163.com}}
\author[4]{Diane Christine Pelejo\thanks{dcpelejo@math.upd.edu.ph}}

\affil[1]{Department of Mathematics, College of William and Mary, Virginia 23187 USA}
\affil[2]{College of Mathematics and Computer Science, Fuzhou University, Fuzhou,
350108, China}
\affil[3]{College of Arts and Sciences, Shanghai Maritime University, Shanghai 201306, China}
\affil[4]{Institute of Mathematics,  College of Science, University of the Philippines Diliman, 1101 Philippines}

\date{}
\begin{document}
\openup 1 \jot
\maketitle

\begin{abstract}
We consider a  homogeneous system of linear equations
of the form $A_\alpha^{\otimes N} \bx =  0$ arising
from the  distinguishability of two quantum operations by $N$
uses in parallel, where the coefficient matrix $A_\alpha$ depends on
a real parameter $\alpha$.
It was conjectured by Duan et al. that the system has
a non-trivial nonnegative
solution if and only if $\alpha$ lies in a certain interval $R_N$
depending on $N$.
We affirm the necessity part of the conjecture and establish the
sufficiency of the conjecture
for $N\leq 10$ by presenting explicit non-trivial nonnegative
solutions for the linear system.
\end{abstract}



\section{Introduction}
Let $M_{m,n}$ (respectively, $M_n$) be the set of $m\times n$
(respectively, $n\times n$)  complex matrices.  Denote by  $H_n$ the set of $n\times n$ Hermitian matrices and by $D_n$ the set of $n\times n$ density matrices, which are positive semidefinite matrices with trace one.

In the mathematical framework of quantum mechanics,
density matrices are used to describe the state of a quantum system.
Quantum operations \cite{nc,Watrous} are trace-preserving,
completely-positive
linear maps from $M_n$ to $M_m$.  It is known \cite{choi,kraus}
that for a quantum operation $\mathcal{E}: M_n\longrightarrow M_m$,
there exists a set of matrices
$\{E_1,\ldots, E_{{n_0}}\}\subset M_{m,n}$, called a set of
Choi-Kraus operators of $\mathcal{E}$, such that
\[\sum\limits_{j=1}^{n_0}E_j^{*}E_j=I_n \quad\mbox{ and }\quad
\mathcal{E}(X)=
\sum\limits_{j=1}^{n_0}E_jXE_j^{*} \quad \mbox{ for any }X\in M_n.\]
For example, the identity map on $M_{\ell}$, denoted by $\mathcal{I}_{\ell}$,
has $\{I_{\ell}\}$ as Choi-Kraus operator.

Two quantum operations  $\mathcal{E}:M_n\longrightarrow M_m$
and $\mathcal{F}:M_n\longrightarrow M_m$, with Choi-Kraus operators
given by $\{E_j\}_{j=1}^{n_0}$ and $\{F_k\}_{k=1}^{n_1}$ are
\textbf{distinguishable by $N$ uses in parallel}
if for some integers $\ell,r$, there exists a nonzero vector
$\mathbf{x} \in
\mathbb{C}^{\ell^{r}\cdot n^{N}}$ such that
\[Y_1=(\mathcal{I}_{\ell}^{\otimes r}\otimes \mathcal{E}^{\otimes N})(\mathbf{x}
\mathbf{x}^*)=\sum_{j_1,\ldots, j_N\in \{1,\ldots,n_0\}} (I_{\ell}^{\otimes r}\otimes E_{j_1}\otimes \cdots \otimes E_{j_N})
\mathbf{x}\mathbf{x}^*
(I_{\ell}^{\otimes r}\otimes E^*_{j_1}\otimes \cdots \otimes E^*_{j_N}) \]
and
\[Y_2=(\mathcal{I}_{\ell}^{\otimes r}\otimes \mathcal{F}^{\otimes N})(\mathbf{x}
\mathbf{x}^*)=\sum_{k_1,\ldots, k_N\in\{1,\ldots, n_1\}} (I_{\ell}^{\otimes r}\otimes F_{k_1}\otimes \cdots \otimes F_{k_N})
\mathbf{x}\mathbf{x}^*
(I_{\ell}^{\otimes r}\otimes F^*_{k_1}\otimes \cdots \otimes F^*_{k_N}) \]
are orthogonal, that is $\mbox{Tr}(Y_1^{*}Y_2)=0$.
One may see \cite{Detal} and its references for the background of
the concept.  In particular, the following results were obtained in
\cite[Theorems 1 and 2]{Detal}.

\begin{proposition} \label{prop1}
Let $\mathcal{E}$ and $\mathcal{F}$ be two quantum operations  with Choi-Kraus operators $\{E_j\}_{j=1}^{n_0}$ and $\{F_k\}_{k=1}^{n_1}$, respectively. Then $\mathcal{E}$ and $\mathcal{F}$ can be perfectly distinguished by $N$ uses in parallel if and only if there exists a density matrix $\rho\in (S^{\otimes N}_{\mathcal{E},\mathcal{F}})^{\perp}$, where
\[S_{\mathcal{E},\mathcal{F}}=\Span\{E_j^*F_k\ |
\ 1\leq j\leq n_0, 1\leq k\leq n_1\} \quad \hbox{ and } \quad
S^{\otimes N}_{\mathcal{E},\mathcal{F}} = \Span\{R^{\otimes N}: R \in
S_{\mathcal{E},\mathcal{F}}\}.\]
\end{proposition}

\begin{proposition}\label{1.2}
Any non-empty subset $T\subseteq M_n$ can be
realized as a spanning set of $S_{\mathcal{E},\mathcal{F}}$ of some pair of
quantum operations $\mathcal{E},\mathcal{F}$.
\end{proposition}

Here we give a short proof of Proposition \ref{1.2}:
Suppose $\Span T$ has a basis $\{A_1, \dots, A_m\} \subseteq M_n$.
Consider the block diagonal matrix $\bA = A_1 \oplus \cdots \oplus A_m$.
If $\bA$
has rank $\tilde k$, then  $\bA = [B_1 \cdots B_m]^*[C_1 \cdots C_m]$,
where $B_1, \dots, B_m, C_1, \dots, C_m$ are $k \times n$ matrices
with $k = \max\{\tilde k,n\}$.
Let $M > 0$ be such that $I_n - \frac{1}{M} \sum_{j=1}^m B_j^*B_j
= B_{m+1}^*B_{m+1}$
and $I_n - \frac{1}{M}\sum_{j=1}^m C_j^*C_j =
C_{m+1}^*C_{m+1}$ for some $k\times n$
matrices $B_{m+1}, C_{m+1}$.
Let $E_1, \dots, E_{m+1}, F_1, \dots, F_{m+1} \in M_{3k,n}$
be such that
$$E_j^* = \frac{1}{\sqrt{M}}[B_j^* | 0_{n,2k}],
\quad F_j^* = \frac{1}{\sqrt{M}}[C_j^* | 0_{n,2k}],
\qquad j = 1, \dots, m,$$
$$E_{m+1}^* = [0_{n,k} | B_{m+1}^* | 0_{n,k} ], \quad
F_{m+1}^* = [0_{n,2 k} | C_{m+1}^*].$$
Then $\sum_{j=1}^{m+1} E_j^*E_j = \sum_{j=1}^{m+1} F_j^*F_j = I_n$, and
$$[E_1 \cdots E_{m+1}]^* [F_1 \cdots F_{m+1}]
= \frac{1}{M}(A_1 \oplus \cdots \oplus A_m \oplus 0_n).$$
If the quantum channels from $M_n$ to $M_{3k}$ have the sets
of Choi-Kraus operators $\{E_1, \dots, E_{m+1}\}$ and
$\{F_1, \dots, F_{m+1}\}$, then
$\Span \{E_i^*F_j: 1 \le i, j \le m+1\}= \Span\{A_1,\dots, A_m\}$. \qed

In \cite{Detal}, the authors considered the quantum channels
${\mathcal E}$ and ${\mathcal F}$
with  $S_{\mathcal{E},\mathcal{F}}$ equal to the span of the set
$$
T_{\alpha}=\left\{\begin{bmatrix}
1 & 0 & 0 \\
0 & e^{i\alpha}& 0\\
0 & 0 & 0
\end{bmatrix}, \
\begin{bmatrix}
0 & 0 & 0 \\
0 & 1  & 0 \\
0 & 0 &  e^{i\alpha} \end{bmatrix}  \right\},
\qquad  \alpha \in [0, 2\pi). $$
It is easy to see that the following conditions
for a density operator $\rho = (\rho_{ij})\in M_{3^N}$ are equivalent.
\begin{itemize}
\item[(a)] The density operator $\rho \in (\Span(T_{\alpha})^{\otimes N})^{\perp}$.
\item[
(b)] The diagonal density operator
$\hat\rho = \diag(\rho_{11}, \dots, \rho_{3^N,3^N})
\in (\Span(T_{\alpha})^{\otimes N})^{\perp}$.
\item[
(c)] The vector $\mathbf{x} = (\rho_{11}, \dots, \rho_{3^N,3^N})^t
\in \mathbb{C}^{3^N}$ satisfies
the homogeneous equation
\begin{equation}\label{maineq}
A_{\alpha}^{\otimes N}\mathbf{x}=0
\qquad \hbox{ with } \quad
A_{\alpha}=\begin{bmatrix}
1 & e^{i\alpha} & 0  \\
0 & 1  & e^{i\alpha}
\end{bmatrix}.
\end{equation}
\end{itemize}

By the above fact, one can focus on finding a non-trivial nonnegative vector
$\mathbf{x} \in \mathbb{C}^{3^N}$ satisfying (\ref{maineq}).
Furthermore, the following remarks and conjecture were made in \cite{Detal}.

\begin{remark}  \rm
If $\alpha\in [0,\frac{\pi}{2})$,
the space $\mbox{Span}(T_{\alpha})$ contains a positive operator and in this case
$(\mbox{Span}(T_{\alpha})^{\otimes N})^{\perp}$ does not contain a density matrix for any positive integer $N$. This makes the corresponding pair of
quantum operations $\mathcal{E},\mathcal{F}$,
satisfying $S_{\mathcal{E},\mathcal{F}}=\mbox{Span}(T)$, indistinguishable. By taking the complex conjugate of equation (\ref{maineq}), we see that there is a
non-trivial nonnegative solution to $A_{\alpha}^{N}\mathbf{x}=0$ if and only if
there is a non-trivial nonnegative solution to $A_{-\alpha}^{N}\mathbf{x}=A_{2\pi-\alpha}^{N}\mathbf{x}=0$. Hence, we only need to focus on the case when $\alpha\in\left[\frac{\pi}{2},\pi\right]$.

\end{remark}

\begin{conjecture} \label{con}
Let $\alpha\in\left[\frac{\pi}{2},\pi\right]$.
The equation {\rm (\ref{maineq})}
has a non-trivial nonnegative solution if and only if $\alpha\in  [\frac{\pi}{2}+\frac{\pi}{2N},\pi]$.
\end{conjecture}

\medskip
In \cite{Detal}, the authors gave
explicit solutions of the equation (\ref{maineq}) for $N\leq 4$. Furthermore,
in Section IV of the paper, it was shown that one may reduce the
complexity of the equation (\ref{maineq}) by finding solution with some
symmetries imposed on its entries, and reduce the
equation to another equation $C_{\alpha,N} \mathbf{y}=0$,
where $C_{\alpha, N}$ is an $(N+1)\times \dfrac{(N+1)(N+2)}{2}$
matrix with full column rank. In Section 2, we will set up the
system $C_{\alpha,N}\by = 0$ and obtain another symmetry for the
solution.
In Section 3, we prove the necessity part of Conjecture \ref{con},
that is, if  $\alpha\in [\frac{\pi}{2},\pi]$ and equation (\ref{maineq})
has a non-trivial nonnegative solution, then
$\alpha\in  [\frac{\pi}{2}+\frac{\pi}{2N},\pi]$.
In Section 4,
we present explicit non-trivial nonnegative solutions $(\ref{maineq})$ for
$\alpha\in  [\frac{\pi}{2}+\frac{\pi}{2N},\pi]$ and $N\leq 10$.
In Section 5,
we provide some additional remarks that may help
in studying the sufficiency part of the conjecture.

\section{A reduction of the linear system}

First, we label the entries of a vector $\mathbf{x}\in \mathbb{C}^{3^N}$ using ternary numbers. That is, we use the ternary number $(j_0,\ldots, j_{N-1})\in \{0,1,2\}^{N}$, for the $j^{th}$ entry of $\mathbf{x}$ when
\[j=1+\sum\limits_{p=0}^{N-1}j_p3^{N-1-p}.\]
For example, we will label the entries of $\mathbf{x}\in \mathbb{C}^{3^2}$ with
$00,01,02,10,11,12,20,21,22$. In the same manner, we label the columns of
$A_{\alpha}^{\otimes N}$ using ternary numbers. Meanwhile, we label the rows of
$A_{\alpha}^{\otimes N}$ using binary numbers.

In \cite[Section IV]{Detal}, it was shown that one may reduce the
complexity of the equation (\ref{maineq}) by finding solution
${\bf x} = [x_J]_{J \in \{0,1,2\}^N} \in \IC^{3^N}$
with entries labeled by $J \in \{0,1,2\}^N$
such that $x_J = x_{\hat J}$ whenever $J = P\hat J$ for a permutation matrix
$P \in M_N$, i.e., the ternary sequences $J$ and $\hat J$  have the same
numbers of $0, 1, 2$ terms. We summarize the result in the following.

\begin{proposition}\label{sym1}
If there is a non-trivial nonnegative solution $\mathbf{x}$ satisfying equation
{\rm (\ref{maineq})}, then there is a non-trivial nonnegative solution
$\hat{\mathbf{x}}=[\hat{x}_J]_{J\in\{0,1,2\}^N}$ such that
\[\hat{x}_{j_0,\ldots, j_{N-1}}=\hat{x}_{k_0,\ldots, k_{N-1}}\]
whenever there exists a permutation $\sigma \in S_{N}$ such that
\[(j_{\sigma(0)},\ldots, j_{\sigma(N-1)})=(k_0,\ldots, k_{N-1}).\]
\end{proposition}

For a triple $(N_0,N_1,N_2)$ of nonnegative integers with $N_0+N_1+N_2=N$,  define the set
\begin{equation}\label{012lab}
[N_0,N_1,N_2]=\{(j_0,j_1,\cdots, j_{N-1})\in \{0,1,2\}^N:
N_{\ell}=\#\{p: j_p=\ell\} \mbox{ for all } \ell \in \{0,1,2\}\},
\end{equation}
of all ternary labels of length $N$ that contains $N_0$ digits equal to $0$, $N_1$ digits equal to $1$ and $N_2$ digits equal to $2$. For example, when $N=2$,
\[ [1,1,0]=\{01,10\}, \qquad [0,1,1]=\{12,21\}, \qquad [2,0,0]=\{00\}. \]
Proposition \ref{sym1} states that if there is a non-trivial nonnegative solution
to $A_{\alpha}^{N}\mathbf{x}=0$, then there is a non-trivial nonnegative solution
$\hat{\mathbf{x}}=[a_J]_{J\in \{0,1,2\}^N}$ such that $a_J=a_K$ whenever
$J,K\in [N_0,N_1,N_2]$. Using this symmetry, $\hat{\mathbf{x}}$ has at most
$p_N$ distinct entries, where $p_N$ is the number of nonnegative integer triples
$(N_0,N_1,N_2)$ satisfying $N_0+N_1+N_2=N$. The total number of such
triples equals the sum of solutions of $N_0 + N_1 = k$ for $k = 0, \dots, N$,
and hence
\begin{equation}\label{hat1}
p_N= 1 + \cdots + (N+1) = \dfrac{(N+1)(N+2)}{2}.
\end{equation}
For example, when $N=2$, we see from equation (\ref{hat1}), that $\hat{x}$ has
at most $p_2=6$ distinct entries. In fact, we may assume that the solution
has the form:
\begin{equation}\label{hat2}
\hat{\mathbf{x}}^T = \begin{bmatrix}
x_{00} & x_{01} &  x_{02} &  x_{10} &  x_{11}
& x_{12} & x_{20} &  x_{21} & x_{22}\end{bmatrix} =  \begin{bmatrix}
a & b & c & b & d & e & c & e & f
\end{bmatrix}.
\end{equation}

In the following, it is convenient to replace
$A_\alpha$ by the matrix
\begin{equation}\label{A}
A_{\alpha}=\begin{bmatrix}
1 & 0 & -e^{2i\alpha}  \\
0 & 1  & e^{i\alpha}
\end{bmatrix}=
\begin{bmatrix}
1 & -e^{i\alpha}\\
0 & 1
\end{bmatrix} \begin{bmatrix}
1 & e^{i\alpha} & 0  \\
0 & 1  & e^{i\alpha}
\end{bmatrix}
= \begin{bmatrix}
1 & 0 & -z^2  \\
0 & 1  &  z
\end{bmatrix} \qquad \hbox{ with } z = e^{i\alpha}.
\end{equation}
Now, let us define a $3^N\times p_N$ matrix $Q_N$ by labeling its rows by ternary numbers in the usual order and labeling its first $N+1$ columns by $[N,0,0],[N-1,0,1],\ldots, [1,0,N-1],[0,0,N]$, then its next $N$ columns by $[N-1,1,0],[N-2,1,1],\ldots, [0,1,N-1]$ and so on; then setting the $(i,j)^{th}$ entry of $Q_N$ equal to $1$ precisely when the ternary label of the $i^{th}$ row is an element of the $j^{th}$ column label as defined in equation (\ref{012lab}). We can then define the following $2^{N}\times p_N$ matrix
\begin{equation}\label{B}
B_{\alpha,N}=A_{\alpha}^{\otimes N} Q_N.
\end{equation}
Notice that for a non-trivial nonnegative solution $\hat{\mathbf{x}}$ satisfying the symmetry described in Proposition \ref{sym1}, we have
\[A_{\alpha}^{\otimes N} \hat{\mathbf{x}}=A_{\alpha}^{\otimes N} Q_N \mathbf{y}=B_{\alpha,N} \mathbf{y}\]
for some nonzero nonnegative vector  $\mathbf{y}\in\mathbb{C}^{p_N}$. Observe $B_{\alpha,N}$ for $N=2,3$ given below,
\[B_{\alpha,2}=\bordermatrix[{[]}]{~ & \mbox{\footnotesize{[200]}} & \mf{[101]} & \mf{[002]} & \mf{[110]} & \mf{[011]} & \mf{[020]}\cr
 \mf{[00]} & 1 & -2 z^2 & z^4 & 0 & 0 & 0 \cr
 \mf{[01]} & 0 & z & -z^3 & 1 & -z^2 & 0 \cr
 \mf{[10]} & 0 & z & -z^3 & 1 & -z^2 & 0 \cr
 \mf{[11]} & 0 & 0 & z^2 & 0 & 2 z & 1},\]
\[B_{\alpha,3}=\bordermatrix[{[]}]{~ & \mf{[300]} & \mf{[201]} & \mf{[102]} &
\mf{[003]} & \mf{[210]} & \mf{[111]} & \mf{[012]} & \mf{[120]} & \mf{[021]} &
\mf{[030]}\cr
\mf{[000]} & 1 & -3 z^2 & 3 z^4 & -z^6 & 0 & 0 & 0 & 0 & 0 & 0 \cr
\mf{[001]}&  0 & z & -2 z^3 & z^5 & 1 & -2 z^2 & z^4 & 0 & 0 & 0 \cr
\mf{[010]}&  0 & z & -2 z^3 & z^5 & 1 & -2 z^2 & z^4 & 0 & 0 & 0 \cr
\mf{[011]} & 0 & 0 & z^2 & -z^4 & 0 & 2 z & -2 z^3 & 1 & -z^2 & 0 \cr
\mf{[100]} &  0 & z & -2 z^3 & z^5 & 1 & -2 z^2 & z^4 & 0 & 0 & 0 \cr
\mf{[101]} & 0 & 0 & z^2 & -z^4 & 0 & 2 z & -2 z^3 & 1 & -z^2 & 0 \cr
\mf{[110]}& 0 & 0 & z^2 & -z^4 & 0 & 2 z & -2 z^3 & 1 & -z^2 & 0 \cr
\mf{[111]} & 0 & 0 & 0 & z^3 & 0 & 0 & 3 z^2 & 0 & 3 z & 1 \cr
}.\]

\begin{proposition}
Let $B_{\alpha,N}$ be defined as in equation {\rm (\ref{B})}, then
\begin{enumerate}
\item[{\rm (a)}] $\mbox{\rm rank}(B_{\alpha,N})=N+1$.
\item[{\rm (b)}] The first $N+1$ columns of $B_{\alpha,N}$ are linearly independent.
\item[{\rm (c)}] If the digits of the binary labels of rows $J$ and $K$
have the same number of zeros (equivalently, the same number of ones),
then the $J^{th}$ and $K^{th}$ rows of $B_{\alpha,N}$ are identical.
\item[{\rm(d)}] Let $J=00\ldots\underbrace{11\ldots 1}_{j}$. Then
\begin{equation}\label{eqn: B_Jfinal}
(B_{\alpha,N})_{J;[N_0,N_1,N_2]}=\binom{N-j}{N_0}\cdot(-z^2)^{N-j-N_0} \cdot \binom{j}{N_1} z^{j-N_1},
\end{equation} where we agree that $\binom{n}{k}=0$ whenever $k>n$.
\end{enumerate}
\end{proposition}
\textit{Proof:}
Let $A_{\alpha}$ be  defined as in equation (\ref{A}).
Denote its entries by $a_{j,\ell}$ where $j\in\{0,1\}$
and $\ell\in\{0,1,2\}$. One can check that if
$J=(j_1,\ldots,j_N)\in \{0,1\}^N$ and
$L=(\ell_1,\ldots,\ell_N)\in \{0,1,2\}^N$, then the $(J,L)$ entry of
$A_{\alpha}^{\otimes N}$ is $\prod_{s=1}^N a_{j_s,\ell_s}$.

We first prove (c). Since $J$ and $K$ have the same number of zeros, there exists $\sigma\in S_N$ such that $J=\sigma(K)$.

Let $N_0,N_1,N_2$ be nonnegative integers with $N_0+N_1+N_2=N$, $\tau\in S_N$, write \[
\tau([N_0,N_1,N_2]):=\{\tau(L)|L\in [N_0,N_1,N_2]\}.
\]
It is easy to verify that $\tau([N_0,N_1,N_2])=[N_0,N_1,N_2]$  for any $\tau$. Then
\begin{eqnarray*}
	(B_{\alpha,N})_{J;[N_0,N_1,N_2]}& = & \sum_{L=\ell_1\ell_2\ldots \ell_{N}\in [N_0,N_1,N_2]} (A_{\alpha}^{\otimes N})_{J,L}\\
	& = & \sum_{L=\ell_1\ell_2\ldots \ell_{N}\in [N_0,N_1,N_2]}\prod_{s=1}^N(A_{\alpha})_{j_s,\ell_s}\\
	& = & \sum_{\sigma(L)\in [N_0,N_1,N_2] } \prod_{s=1}^N (A_{\alpha})_{j_{\sigma(s)},\ell_{\sigma(s)}}\\
	& = & \sum_{L\in [N_0,N_1,N_2] } \prod_{s=1}^N (A_{\alpha})_{k_s,\ell_s}\\
	& = & (B_{\alpha,N})_{K;[N_0,N_1,N_2]},
\end{eqnarray*}
where $J=j_1j_2\ldots j_N$ and $K=k_1k_2\ldots k_N$.
Thus the $J^{th}$ row and $K^{th}$ rows of $B_{\alpha,N}$ are identical.

Note that \begin{eqnarray}\label{eqn: B_Jbasic}
(B_{\alpha,N})_{J;[N_0,N_1,N_2]}=\sum_{L=\ell_1\ell_2\ldots\ell_N\in [N_0,N_1,N_2]}\prod_{s=1}^{N-j}a_{0,\ell_s}\prod_{t={N-j+1}}^N a_{1,\ell_t}.
\end{eqnarray}
Let $L\in [N_0,N_1,N_2]$ corresponding to a nonzero term in the formula (\ref{eqn: B_Jbasic}). Since $a_{10}=0$, then $\{s|\ell_s=0\}\subseteq [N-j]$. Since $\#\{s|\ell_s=0\}=N_0$, there are $\binom{N-j}{N_0}$ different choices for the positions of 0s in $L$. Now suppose that the positions of $0$s have been chosen, then for $s\in[N-j]\backslash\{s|\ell_s=0\}$, $\ell_s$ can't be 1 since $a_{01}=0$. Thus there are $N-j-N_0$ terms of $-z^2$ in the first product.

For the second product, $\ell_{N-j+1}\cdots\ell_{N}$ must contain all the $N_1$ 1s in $L$ since $a_{01}=0$, and there are $\binom{j}{N_1}$ different choices. After the positions of 1s been chosen, all the $\ell_t$ left must be 2, thus the corresponding terms in the product are $z$, yielding that the formula (\ref{eqn: B_Jfinal}) holds.

By formula (\ref{eqn: B_Jfinal}), the submatrix of $B_{\alpha,N}$ obtained by taking only the rows labeled by
$00\cdots 00$, $00\cdots 01,\ldots, 11\cdots 11$, and columns labeled by $[N,0,0]$, $[N-1,0,1],\ldots,[1,0,N-1]$, $[0,0,N]$, is an
upper triangular matrix with nonzero diagonals. Thus (a) and (b) hold.
\qed

Now, define the $(N+1) \times p_N$ matrix  $C_{\alpha,N}$ as the submatrix of $B_{\alpha,N}$ obtained by taking only the rows labeled by $00\cdots 00$, $00\cdots 01,\ldots, 11\cdots 11$. This makes $C_{\alpha,N}$ a full row rank matrix such that
$B_{\alpha,N}\mathbf{y}=0$ if and only if $C_{\alpha,N}\mathbf{y}=0$. We illustrate $C_{\alpha,2}$ and $C_{\alpha,3}$ below, \[C_{\alpha,2}=\bordermatrix[{[]}]{~ & \mf{[200]} & \mf{[101]} & \mf{[002]} & \mf{[110]} & \mf{[{011}]} & \mf{[020]}\cr
\mf{[00]} & 1 & -2 z^2 & z^4 & 0 & 0 & 0 \cr
 \mf{[01]} & 0 & z & -z^3 & 1 & -z^2 & 0 \cr
 \mf{[11]} & 0 & 0 & z^2 & 0 & 2 z & 1},\]
\[C_{\alpha,3}=\bordermatrix[{[]}]{~ & \mf{[300]} & \mf{[201]} & \mf{[102]} &
\mf{[003]} & \mf{[210]} & \mf{[111]} & \mf{[012]} & \mf{[120]} & \mf{[021]} &
\mf{[030]}\cr
\mf{[000]} & 1 & -3 z^2 & 3 z^4 & -z^6 & 0 & 0 & 0 & 0 & 0 & 0 \cr
\mf{[001]} &  0 & z & -2 z^3 & z^5 & 1 & -2 z^2 & z^4 & 0 & 0 & 0 \cr
\mf{[011]} & 0 & 0 & z^2 & -z^4 & 0 & 2 z & -2 z^3 & 1 & -z^2 & 0 \cr
\mf{[111]} & 0 & 0 & 0 & z^3 & 0 & 0 & 3 z^2 & 0 & 3 z & 1 \cr
}.\]
Notice that we can write $C_{\alpha,N}$ as
\begin{equation}\label{CaN}
C_{\alpha,N}=\begin{bmatrix}
\Gamma_N & \vline & \begin{matrix}
0_{1,N}\\
D_{N,1} \Gamma_{N-1}
\end{matrix}
& \vline &
\begin{matrix}
0_{2,N-1}\\
D_{N,2} \Gamma_{N-2}
\end{matrix}  &\vline &\cdots &\vline &  \begin{matrix}
0_{N,1}\\
D_{N,N}\Gamma_0\
\end{matrix}
\end{bmatrix},
\end{equation}
where for $n = N, \dots, 0,$ and $k \le n$,
$D_{n,k}$ is the diagonal matrix $\mbox{diag}\Big(\binom{k}{k},\binom{k+1}{k},\ldots,\binom{n}{k}\Big)\in M_{n-k+1}$,
and
$$\Gamma_n =\begin{bmatrix} 1 & & & & \cr
& z & & & \cr
& & \ddots &&\cr
& & & \ddots &\cr
& & & & z^n\cr\end{bmatrix}
 \begin{bmatrix}
1 & \binom{n}{1}(-z^2) &  \binom{n}{2}(-z^2)^2& \cdots&
\binom{n}{n} (-z^2)^n \cr
&&&&\cr
0 & 1 & \binom{n-1}{1}(-z^2) &  \cdots& \binom{n-1}{n-1} (-z^2)^{n-1}\cr
&&&&\cr
\vdots& \vdots & \vdots & \vdots &\vdots\cr
&&&&\cr
 0 & 0 & 0 & \cdots & 1\cr\end{bmatrix}.$$
So,
 $\Gamma_n$ is an $(n+1)\times (n+1)$ upper triangular matrix
whose $(j,k)$ entry is given by
\[\left\{\begin{array}{ll}
(-1)^{k-j}\binom{n+1-j}{k-j}z^{j-1+2(k-j)}
& \mbox{ if } k\geq j,\\
0 & \mbox{ if } j>k.
\end{array}\right.  \]

\section{Necessity  of Conjecture \ref{con}}

To prove the necessity of Conjecture \ref{con},
we demonstrate another symmetry one may impose on the solution of
$\bx$ of the equation (\ref{maineq}).

\begin{proposition} \label{sym2}
Suppose $\bx \in \IR^{3^N}$ satisfies equation {\rm (\ref{maineq})},
and $\hat \bx$ is obtained from $\bx$ by exchanging
the entries $\bx_j = \bx_{3^N-j}$ whenever $1 \le j \le (3^N-1)/2$.
Then  $A^{\otimes N} \hat \bx = 0$.
\end{proposition}

Note that if the entries of
$\bx$ and $\hat \bx$ are labeled by $x_{j_1\cdots j_N}$ and $\hat x_{j_1\cdots j_N}$
using ternary sequences  $j_1\cdots j_N\in \{0,1,2\}^N$, then
$x_{j_1\cdots j_N} = \hat x_{(2-j_1)\cdots (2-j_N)}$.
\\
\textit{Proof.}
Let $\tilde{A}_{\alpha}=\begin{bmatrix}
1 & e^{i\alpha} & 0\\
0 & e^{-i\alpha} & 1
\end{bmatrix} $, $J=\begin{bmatrix}
0 & 1 \\ 1 & 0
\end{bmatrix}$ and $K=\begin{bmatrix}
0 & 0 & 1\\
0 & 1 & 0\\
1 & 0 & 0
\end{bmatrix}$.
Then for $A_\alpha$ defined in (\ref{maineq}),
\[\begin{bmatrix}
1 & 0\\
0 & e^{-i\alpha}
\end{bmatrix}
A_{\alpha}=\tilde{A}_{\alpha}=J\overline{\tilde{A}_{\alpha}}K.\]
Thus,
$\mathbf{x}\in {\rm Null}(A_{\alpha}^{\otimes N})$
if and only if $\mathbf{x}\in {\rm Null}(\tilde{A}_{\alpha}^{\otimes N})$.
Additionally, if $\mathbf{x}$ is real,
\[\mathbf{x}\in {\rm Null}(\tilde{A}_{\alpha}^{\otimes N})\quad
\Longleftrightarrow \quad
\tilde{A}_{\alpha}^{\otimes N} \mathbf{x}=\mathbf{0}\quad
\Longleftrightarrow \quad \overline{\tilde{A}_{\alpha}^{\otimes N} }\mathbf{x}
=\mathbf{0} \quad  \Longleftrightarrow \quad K^{\otimes N}\mathbf{x}\in
{\rm Null}(\tilde{A}_{\alpha}^{\otimes N}).\]
So, $\hat x_{j_1 \cdots j_N} = \hat x_{2-j_1 \cdots 2-j_N}$.
Thus, we can assume that $x_{i_1i_2\cdots i_n}=x_{(2-i_1)(2-i_2)\cdots (2-i_N)}$.
\qed

By the above proposition and the discussion  in Section 2, we see that
the system $A_\alpha^{\otimes N} \bx =  0$ has a non-trivial nonnegative
solution if and only if the system $C_{\alpha,N}\mathbf{y}=0$
has a non-trivial nonnegative solution $\by$. We have the following.

\begin{theorem}
Let $\alpha\in [\frac{\pi}{2},\pi]$.
If the equation
$C_{\alpha,N}\mathbf{y}=0$
has a non-trivial nonnegative solution $\by$,
then $\alpha\in \left[\frac{\pi}{2}+\frac{\pi}{2N},\pi\right]$.
\end{theorem}
\textit{Proof.}
 We consider the reduced equation $C_{\alpha,N}\by = 0$
 with $z = e^{i\alpha}$ as shown in Section 2.
 Let
 \begin{eqnarray*}
 \by &=& (y_{[N00]},y_{[(N-1)01]},\ldots,y_{[00N]},y_{[(N-1)10]},\ldots,
 y_{[01(N-1)]},\ldots,\ldots,y_{[0N0]} )^t\\
 &=& (y_{0,0}, y_{0,1}, \dots, y_{0,N}, y_{1,0}, \dots, y_{1,N-1},
 \dots, y_{N,0})^t
\end{eqnarray*}
 be a nonnegative solution
 of $C_{\alpha,N}\by = 0$. We will show that if
 $\alpha \in [\frac{\pi}{2},\frac{\pi}{2}+\frac{\pi}{2N})$, then
 $\by$ is a zero vector, which is a contradiction.

\noindent
{\bf Case 1.} $N$ is even.
 Since $y_{j,k} = y_{j,N-j-k}$,
   we may rewrite the first equation of the linear system as
   \begin{equation}\label{eq: evenfirst}
      \sum_{k = 0}^{N/2-1} (-1)^k \binom{N}{k}(z^{2k}+z^{2N-2k})y_{0,k} + (-1)^{N/2} \binom{N}{N/2}z^{N} y_{0,N/2} = 0.
   \end{equation}
   Divided by $z^N$, (\ref{eq: evenfirst}) reduces to
   \begin{equation}\label{eq: even2}
      \sum_{k=0}^{N/2-1}(-1)^k \binom{N}{k}2\cos((N-2k)\alpha) y_{0,k} +
      (-1)^{N/2} \binom{N}{N/2} y_{0,N/2} = 0.
   \end{equation}

   Let $\theta = \alpha - \frac{\pi}{2} $. Since we assume that $\alpha\in [\frac{\pi}{2},\frac{\pi}{2}+\frac{\pi}{2N})$, then
   $\theta \in [0,\frac{\pi}{2N}) $ so that $\cos(m\theta)$
   are all positive for $1\le m\le N$.

   Now replace $\alpha$ in (\ref{eq: even2}) with $\theta$, we have
   \begin{equation}
      (-1)^{N/2}\left(\sum_{k=0}^{N/2-1} \binom{N}{k}
      2\cos((N-2k)\theta) y_{0,k}+ \binom{N}{N/2} y_{0,N/2}\right)= 0.
   \end{equation}
   Since all the coefficients of $y_{0,k}$ are nonnegative, $y_{0,k} =0$ for all $k=0,1,\ldots,N$.

\noindent
{\bf Case 2.} $N$ is odd.
Since   $y_{j,k} = y_{j,N-j-k}$,
   we may rewrite
the first equation of the linear system as
   \begin{equation}\label{eq: oddfirst}
      \sum_{k = 0}^{(N-1)/2} (-1)^k \binom{N}{k}(z^{2k}-z^{2N-2k})y_{0,k}  = 0.
   \end{equation}
Dividing the equation by $iz^N$, and replacing $\alpha$ with
$\theta= \alpha - \frac{\pi}{2} $, we get
   \begin{equation}
      (-1)^{\frac{N+1}{2}}\left(\sum_{k=0}^{(N-1)/2}\binom{N}{k}
      2\cos((N-2k)\theta)y_{0,k}\right) =0,
   \end{equation}
   by the same reason as the even case, $y_{0,k}$ needs to be $0$ for all
   $k=0,1,\ldots,N$.

   For $y_{1,0},y_{1,1},\ldots,y_{1,N-1}$,  since it is already proved that when
   $\alpha\in [\frac{\pi}{2},\frac{\pi}{2}+\frac{\pi}{2N})$,
$y_{0,0}=y_{0,1}=\cdots=y_{0,N} = 0$, the second equation of
$C_{\alpha,N}\by = 0$ becomes the same as the first equation of
$C_{\alpha,N-1}\by = 0$. By induction on $N$, since
$[\frac{\pi}{2},\frac{\pi}{2}+\frac{\pi}{2N})
\subseteq [\frac{\pi}{2},\frac{\pi}{2}+\frac{\pi}{2(N-1)})$,
we have $y_{1,0}=y_{1,1}=\cdots=y_{1,N-1} = 0$. Furthermore,
by induction on $j$, we have $y_{j,0}=\cdots= y_{j,N-j}=0$ for $j= 0,1,\ldots, N$,
which means $\by = 0$, completing the proof.
\qed

\begin{corollary}
Let $\alpha\in [\frac{\pi}{2},\pi]$.
If the equation $A_{\alpha}^{\otimes N}\mathbf{x}=0$ has a non-trivial
nonnegative solution $\bx$, then
$\alpha\in \left[\frac{\pi}{2}+\frac{\pi}{2N},\pi\right]$.
\end{corollary}

\section{Explicit solution of the system \texorpdfstring{$C_{\alpha,N} \by = 0$}{C{a,N}=0} when
\texorpdfstring{$N\leq 10$}{N<=10}}
Note that if $\alpha\in  [\frac{\pi}{2}+\frac{\pi}{2N},\pi]\subseteq
[\frac{\pi}{2}+\frac{\pi}{2(N+1)},\pi] $ and $\bx\in \mathbb{C}^{3^N}$
satisfy $A_{\alpha}^{\otimes N} \mathbf{x}=0$, then for any nonnegative vector
$\mathbf{y}\in \mathbb{C}^3$, we have
$A_{\alpha}^{\otimes(N+1)}(\mathbf{x}\otimes \mathbf{y})=0$.
Thus, for $N\leq 10$, it is enough to find a non-trivial nonnegative solution to
$C_{\alpha,N}\mathbf{y}=0$  when $\alpha\in
\left[\frac{\pi}{2}+\frac{\pi}{2N},\frac{\pi}{2}+ \frac{\pi}{2(N-1)}\right)$.
In the next lemma, we determine the exact location of $e^{ik\alpha}$ in the
Argand plane. Let $Q_1,Q_2, Q_3, Q_4$ denote the four quadrants of the complex
plane.

\begin{lemma}\label{quadrants}
For $N\geq 2$, let
\begin{equation}\label{alpha}
\theta\in \left[\frac{\pi}{2N},
\quad  \frac{\pi}{2(N-1)}\right) \quad \mbox{ and }
\quad \alpha=\theta+\frac{\pi}{2}.
\end{equation}
Then $N\alpha \in Q_{N+2\pmod{4}}$,
and for $k < N$ we have $k\alpha \in Q_{k+1 \pmod{4}}$.
\end{lemma}
\it Proof. \rm Note that $[a,b]\subseteq Q_r$ if and only if there exists
$\ell$ such that  $r\equiv \ell+1\pmod{4}$ and
$\frac{\ell}{2}\pi\leq a\leq b\leq \frac{\ell+1}{2}\pi$.
Since $\alpha \in \left[\pi\left(\frac{1}{2} +
\frac{1}{2{N}}\right),
\pi\left(\frac{1}{2} + \frac{1}{2({N}-1)}\right)\right)$,
then $N\alpha\in \left[\pi\left(\frac{N+1}{2}\right),
\pi\left(\frac{N+1}{2} + \frac{1}{2(N-1)}\right)\right)$.
Note that $\frac{1}{2(N-1)}\leq \frac{1}{2}$ and hence if $\ell=N+1$,
then $\alpha\in Q_r$ where $r\equiv \ell+1\equiv N+2\pmod{4}$.
On the other hand, if $0\leq k<N$, then $\frac{k}{2(N-1)}\leq \frac{1}{2}$.
Note that $k\alpha\in \left[\pi\left(\frac{k}{2} + \frac{k}{2N}\right),
\pi\left(\frac{k}{2} + \frac{k}{2(N-1)}\right)\right)$. Thus, if $\ell=k$
then $k\alpha\in Q_r$ where $r\equiv \ell+1\equiv k+1\pmod{4}$.\qed

\medskip
We now present some explicit non-trivial nonnegative solutions to $C_{\alpha,N}\mathbf{y}=0$.
One can use the preceding lemma
to verify that the given $\mathbf{y}$ is nonzero and nonnegative.

\begin{enumerate}
\item For $N=1$, we have $\alpha= \pi$ and a non-trivial nonnegative solution given by
\[\mathbf{y}^T= \bordermatrix[{[]}]{~ & \mbox{\footnotesize{[100]}} &
\mf{[001]} & \mf{[010]} \cr
 & 1 & 1 & 1 }.
\]
\item For $N=2$, we have
$\alpha\in \left[\frac{3\pi}{4},\pi\right)$ and a non-trivial nonnegative solution given by
\[\mathbf{y}^T= \bordermatrix[{[]}]{~ & \mbox{\footnotesize{[200]}} & \mf{[101]} & \mf{[002]} & \mf{[110]} & \mf{[011]} & \mf{[020]} \cr
 & 1 & \cos 2\alpha & 1 & -\cos \alpha & -\cos \alpha & 1
 }.\]
\item For $N=3$, we have $\alpha\in \left[ \frac{2\pi}{3}, \frac{3\pi}{4}\right)$ and a non-trivial nonnegative solution given by
\[\mathbf{y}^T=\!\!\! \bordermatrix[{[]}]{~ & \mbox{\footnotesize{[300]}} & \mf{[201]} & \mf{[102]} & \mf{[003]}  & \mf{[210]} & \mf{[111]}
&\mf{[012]} & \mf{[120]} & \mf{[021]} & \mf{[030]}   \cr
& 3\sin\alpha & \sin3\alpha & \sin3\alpha & 3\sin\alpha & -\sin2\alpha
& -\sin2\alpha & -\sin2\alpha &  \sin\alpha & \sin\alpha & 0
 }.\]
\item For $N=4$, we have $\alpha\in
\left[\frac{5\pi}{8},\frac{2\pi}{3}\right)$ and a non-trivial nonnegative solution given by \[\mathbf{y}^T = \begin{bmatrix}
\mathbf{a}_0 & \mathbf{a}_1 & \mathbf{a}_2 & \mathbf{a}_3 & \mathbf{a}_4
\end{bmatrix},\mbox{ where }\]
\[\edef\savedbaselineskip{\the\baselineskip\relax}
\begin{array}{rcl}
\mathbf{a}_0  & = &   {\baselineskip=\savedbaselineskip\bordermatrix[{[]}]{ & \mf{[400]} & \mf{[301]} & \mf{[202]} & \mf{[103]} & \mf{[004]} \cr
 & 6 & 0 & -2\cos4\alpha &  0 & 6 }},
\medskip \\
\mathbf{a}_1  & = &  {\baselineskip=\savedbaselineskip\bordermatrix[{[]}]{~ & \mf{[310]} & \mf{[211]} & \mf{[112]} & \mf{[013]} \cr
&-3\cos\alpha & \cos3\alpha &  \cos3\alpha & -3\cos\alpha
}},
\medskip\\
\begin{bmatrix}
\mathbf{a}_2 & \mathbf{a}_3 & \mathbf{a}_4
\end{bmatrix}  & = &   {\baselineskip=\savedbaselineskip\bordermatrix[{[]}]{~ & \mf{[220]} & \mf{[121]} & \mf{[022]} & \mf{[130]} & \mf{[031]} & \mf{[040]}\cr
& 2 & 0 & 2 &-3\cos\alpha & -3\cos\alpha & 6
}}.
\end{array}
\]
\item For $N=5$, we have $\alpha\in \left[\frac{3\pi}{5}, \frac{5\pi}{8}\right)$ and a non-trivial nonnegative solution given by
\[\mathbf{y}^T = \begin{bmatrix}
\mathbf{a}_0 & \mathbf{a}_1 & \mathbf{a}_2 & \mathbf{a}_3 & \mathbf{a}_4 & \mathbf{a}_5
\end{bmatrix}, \mbox{ where}\]
\[\edef\savedbaselineskip{\the\baselineskip\relax} \begin{array}{rcl}
\mathbf{a}_0  & = &  {\baselineskip=\savedbaselineskip \bordermatrix[{[]}]{ & \mf{[500]} & \mf{[401]} & \mf{[302]} & \mf{[203]} & \mf{[104]} & \mf{[005]} \cr
 & 20\sin \alpha & 0 & -2\sin 5\alpha & -2\sin 5\alpha  & 0 & 20\sin\alpha }},
 \medskip\\
\mathbf{a}_1  & =  &  {\baselineskip=\savedbaselineskip\bordermatrix[{[]}]{ & \mf{[410]} & \mf{[311]} & \mf{[212]} & \mf{[113]} & \mf{[014]} \cr
 & -4\sin 2\alpha & \sin 4\alpha & 2\sin 4\alpha & \sin 4\alpha & -4\sin 2\alpha }},
 \medskip\\
 \mathbf{a}_2 & = &   {\baselineskip=\savedbaselineskip\bordermatrix[{[]}]{ & \mf{[320]} & \mf{[221]} & \mf{[122]} & \mf{[023]} \\
 & 3\sin\alpha & -\sin 3\alpha & -\sin 3\alpha & 3\sin\alpha
 }},
 \medskip\\
 \begin{bmatrix}
\mathbf{a}_3 & \mathbf{a}_4 & \mathbf{a}_5
\end{bmatrix}   & = &   {\baselineskip=\savedbaselineskip\bordermatrix[{[]}]{ & \mf{[230]} & \mf{[131]} & \mf{[032]} & \mf{[140]} & \mf{[041]} & \mf{[ 050]}\cr
& -\sin 2\alpha & 0 & -\sin 2\alpha & 2\sin \alpha & 2\sin \alpha & 0
}}.
\end{array}\]
\item For $N=6$, we have  $\alpha\in \left[\frac{7\pi}{12}, \frac{3\pi}{5}\right)$ and a non-trivial nonnegative solution given by
\[\mathbf{y}^T = \begin{bmatrix}
\mathbf{a}_0 & \mathbf{a}_1 & \mathbf{a}_2 & \mathbf{a}_3 & \mathbf{a}_4 & \mathbf{a}_5 & \mathbf{a}_6
\end{bmatrix}, \mbox{ where}\]
\[\begin{array}{rcl}
\mathbf{a}_0
& = & \left[\begin{array}{ccccccc}
20 & 0 & 0 & 2\cos 6\alpha & 0 & 0& 20
\end{array}\right],
\medskip\\
\mathbf{a}_1
& = & \left[\begin{array}{cccccc}
 -10\cos\alpha & 0 &-\cos5\alpha &  -\cos5\alpha & 0 & -10\cos\alpha
\end{array}\right],
\medskip\\
\begin{bmatrix}
\mathbf{a}_2 & \mathbf{a}_3
\end{bmatrix}
& = & \left[\begin{array}{ccccccccc}
3 & 0 &\cos 4\alpha & 0& 3 & \cos3\alpha & 0 & 0 & \cos3\alpha
\end{array}\right],
\medskip\\
\begin{bmatrix}
\mathbf{a}_4 & \mathbf{a}_5 & \mathbf{a}_6
\end{bmatrix}
& = & \left[\begin{array}{cccccc}
-2\cos2\alpha & 0 & -2\cos2\alpha & 0 & 0 & 5
\end{array}\right].
\end{array}\]
\item For $N=7$, we have $\alpha\in \left[\frac{4\pi}{7}, \frac{7\pi}{12}\right)$ and a non-trivial nonnegative solution given by
\[\mathbf{y}^T = \begin{bmatrix}
\mathbf{a}_0 & \mathbf{a}_1 & \mathbf{a}_2 & \mathbf{a}_3 & \mathbf{a}_4 & \mathbf{a}_5 & \mathbf{a}_6 & \mathbf{a}_7
\end{bmatrix}, \mbox{ where}\]
\[\begin{array}{rcl}
\mathbf{a}_0
 & = & \left[\begin{array}{cccccccc}
140\sin\alpha &0 & 0 & 4\sin7\alpha & 4\sin7\alpha & 0& 0 & 140\sin\alpha
\end{array}\right],
\medskip\\
\mathbf{a}_1 & = & \left[\begin{array}{ccccccc}
-30\sin2\alpha & 0 & -2\sin6\alpha & -4\sin6\alpha & -2\sin6\alpha & 0 & -30\sin2\alpha
\end{array}\right],
\medskip\\
\mathbf{a}_2 & = & \left[\begin{array}{cccccc}
20\sin\alpha & 0 & 2\sin5\alpha & 2\sin5\alpha & 0 & 20\sin\alpha
\end{array}\right],
\medskip\\
\mathbf{a}_3 & = & \left[\begin{array}{ccccc}
 2\sin4\alpha-4\sin2\alpha & \sin4\alpha & 0 & \sin4\alpha & 2\sin4\alpha-4\sin2\alpha
\end{array}\right],
\medskip\\
\mathbf{a}_4 & = & \left[\begin{array}{cccc}
-4\sin3\alpha+6\sin\alpha & -2\sin3\alpha & -2\sin3\alpha & -4\sin3\alpha+6\sin\alpha
\end{array}\right],
\medskip\\
\begin{bmatrix}
\mathbf{a}_5 & \mathbf{a}_6 & \mathbf{a}_7
\end{bmatrix}
& = & \left[\begin{array}{cccccc}
 0 & 0 & 0 & 10\sin\alpha & 10\sin\alpha & 0
\end{array}\right].
\end{array} \]
\item For $N=8$, we have $\alpha\in \left[\frac{9\pi}{16}, \frac{4\pi}{7}\right)$ and a non-trivial nonnegative solution given by
\[\mathbf{y}^T = \begin{bmatrix}
\mathbf{a}_0 & \mathbf{a}_1 & \mathbf{a}_2 & \mathbf{a}_3 & \mathbf{a}_4 & \mathbf{a}_5 & \mathbf{a}_6 & \mathbf{a}_7 & \mathbf{a}_8
\end{bmatrix}, \mbox{ where}\]
\[\begin{array}{rcl}
\mathbf{a}_0
 & = & \left[\begin{array}{ccccccccc}
140 & 0 & 0 & 0 &-4\cos8\alpha & 0& 0& 0& 140
\end{array}\right],
\medskip\\
\mathbf{a}_1
& = & \left[\begin{array}{cccccccc}
 -70\cos\alpha & 0 & 0 &2\cos7\alpha & 2\cos7\alpha & 0& 0& -70\cos\alpha
\end{array}\right],
\medskip\\
\mathbf{a}_2
& = & \left[\begin{array}{ccccccc}
  20 & 0 & 0 & -2\cos6\alpha & 0& 0& 20
\end{array}\right],
\medskip\\
\mathbf{a}_3 &
= & \left[\begin{array}{cccccc}
5\cos3\alpha  &-\cos5\alpha & 0 & 0 & -\cos5\alpha& 5\cos3\alpha
\end{array}\right],
\medskip\\
\mathbf{a}_4
& = & \left[\begin{array}{ccccc}
-8\cos2\alpha & 2\cos4\alpha & 0 & 2\cos4\alpha & -8\cos2\alpha
\end{array}\right],
\medskip\\
\begin{bmatrix}
\mathbf{a}_5 & \mathbf{a}_6 & \mathbf{a}_7 & \mathbf{a}_8
\end{bmatrix}
& = & \left[\begin{array}{cccccccccc}
0 & 0 & 0 & 0  & 10 & 0 & 10 & -35\cos\alpha & -35\cos\alpha & 140
\end{array}\right].
\end{array} \]
\item
For $N=9$, we have $\alpha\in \left[\frac{5\pi}{9}, \frac{9\pi}{16}\right)$ and a non-trivial nonnegative solution given by
\[\mathbf{y}^T= \begin{bmatrix}
\mathbf{a}_0 & \mathbf{a}_1 & \mathbf{a}_2 & \mathbf{a}_3 & \mathbf{a}_4 & \mathbf{a}_5 & \mathbf{a}_6 & \mathbf{a}_7 & \mathbf{a}_8 & \mathbf{a}_9
\end{bmatrix}, \mbox{ where}\]
\[\begin{array}{rcl}
\mathbf{a}_0
& = & \left[\begin{array}{cccccccccc}
504\sin\alpha & 0 & 0 & 0 & -4\sin9\alpha & -4\sin9\alpha & 0& 0& 0& 504\sin\alpha
\end{array}\right],
\medskip\\
\mathbf{a}_1
& = & \left[\begin{array}{ccccccccc}
-112\sin2\alpha & 0 & 0 & 2\sin8\alpha & 4\sin8\alpha & 2\sin8\alpha& 0 & 0 & -112\sin2\alpha
\end{array}\right],
\medskip\\
\mathbf{a}_2
& = & \left[\begin{array}{cccccccc}
70\sin\alpha & 0 & 0 & -2\sin7\alpha & -2\sin7\alpha & 0 & 0 & 70\sin\alpha
\end{array}\right],
\medskip\\
\mathbf{a}_3
& = & \left[\begin{array}{ccccccc}
6\sin4\alpha-15\sin2\alpha & -\sin6\alpha & -\sin6\alpha & 0 & -\sin6\alpha  & -\sin6\alpha & 6\sin4\alpha-15\sin2\alpha
\end{array}\right],
\medskip\\
\mathbf{a}_4
& = & \left[\begin{array}{cccccc}
-10\sin3\alpha+20\sin\alpha & 2\sin5\alpha & 2\sin5\alpha & 2\sin5\alpha & 2\sin5\alpha& -10\sin3\alpha\!+\!20\sin\alpha
\end{array}\right],
\medskip\\
\mathbf{a}_5 & = & \left[\begin{array}{ccccc}
4\sin4\alpha & 0 & 0 & 0 & 4\sin4\alpha
\end{array}\right],
\medskip\\
\mathbf{a}_6 & = & \left[\begin{array}{cccc}
 15\sin\alpha-12\sin3\alpha & -5\sin3\alpha & -5\sin3\alpha & 15\sin\alpha-12\sin3\alpha
\end{array}\right],
\end{array}\]
\[\begin{array}{rcl}
\begin{bmatrix}
 \mathbf{a}_7 & \mathbf{a}_8 & \mathbf{a}_9
\end{bmatrix} & = & \left[\begin{array}{cccccc}
0 & 0 & 0 & 56\sin\alpha & 56\sin\alpha & 0
\end{array}\right]. \phantom{this text is for extra extra extra spacE}
\end{array} \]
\item For $N=10$, we have $\alpha\in \left[\frac{11\pi}{20}, \frac{5\pi}{9}\right)$ and a non-trivial nonnegative solution is given by
\[\mathbf{y}^T = \left[\begin{array}{ccccccccccc}
\mathbf{a}_0 & \mathbf{a}_1 & \mathbf{a}_2 & \mathbf{a}_3 & \mathbf{a}_4 & \mathbf{a}_5 & \mathbf{a}_6 & \mathbf{a}_7 & \mathbf{a}_8 & \mathbf{a}_9 & \mathbf{a}_{10}
\end{array}\right], \mbox{ where}\]
\[\begin{array}{rcl}
\mathbf{a}_0 & = & \left[\begin{array}{ccccccccccc}
504 & 0 & 0 & 0 & 0 & 4\cos10\alpha & 0 & 0& 0 & 0 & 504
\end{array}\right],
\medskip\\
\mathbf{a}_1 & = & \left[\begin{array}{cccccccccc}
-252\cos\alpha & 0 & 0 & 0 & -2\cos9\alpha & -2\cos9\alpha& 0 & 0 & 0& -252\cos\alpha
\end{array}\right],
\medskip\\
\mathbf{a}_2 & = & \left[\begin{array}{ccccccccc}
70 & 0 & 0 & 0 & 2\cos8\alpha & 0 & 0 & 0 & 70
\end{array}\right],
\medskip\\
\mathbf{a}_3 & = & \left[\begin{array}{cccccccc}
21\cos3\alpha & 0 & \cos7\alpha & 0 & 0 & \cos7\alpha & 0 & 21\cos3\alpha
\end{array}\right],
\medskip\\
\mathbf{a}_4 & = & \left[\begin{array}{ccccccc}
-30\cos2\alpha & 0 & -2\cos6\alpha & 0 & -2\cos6\alpha & 0& -30\cos2\alpha
\end{array}\right],
\medskip\\
\mathbf{a}_5 & = & \left[\begin{array}{cccccc}
-4\cos5\alpha & 0 & 0 & 0 & 0 & -4\cos5\alpha
\end{array}\right],
\medskip\\
\mathbf{a}_6 & = & \left[\begin{array}{ccccc}
12\cos4\alpha+15 & 0 & 5\cos4\alpha & 0 & 12\cos4\alpha+15
\end{array}\right],
\end{array}\]
\[\begin{array}{rcl}
\begin{bmatrix}
 \mathbf{a}_7 & \mathbf{a}_8 & \mathbf{a}_9 & \mathbf{a}_{10}
\end{bmatrix} & = & \left[\begin{array}{cccccccccc} 0 & 0 & 0 & 0 & -56\cos2\alpha & 0 -56\cos2\alpha & 0 & 0 & 504 \end{array}\right].
\end{array} \]
\end{enumerate}

\section{Final Remark}

It would be nice to affirm the sufficiency of the Conjecture 1.4 for $N > 10$.
Ideally, one can describe a non-trivial nonnegative
solution of the linear system for every positive
integer $N$. One may also consider finding an existence proof.
In this connection, we have the following proposition.
We will continue to use the notation $C_{\alpha,N}$ and consider the
reduced system $C_{\alpha,N}\by = 0$.

\begin{proposition} Suppose
$\alpha\in \left[\frac{\pi}{2}+\frac{\pi}{2N},\pi\right]$.
The following conditions are equivalent.
\begin{itemize}
\item[{\rm (a)}]
The system $C_{\alpha,N} \by  = 0$ has no non-trivial nonnegative solution.
\item[{\rm (b)}]
There is a complex vector $\bu = (\xi_0, \dots, \xi_N)$
with all entries having positive real parts
such that
all the entries of $\bu C_{\alpha,N}$
has positive real parts.
\end{itemize}
\end{proposition}

\it Proof. \rm We convert the system  $C_{\alpha, N}\by = 0$
to a real linear system
\begin{equation}
\label{realsys}
\tilde C_{\alpha,N} \by = 0, \qquad \hbox{ where } \quad
\tilde C_{\alpha,N} = \begin{bmatrix} \Re(C_{\alpha,N}) \cr
\Im(C_{\alpha,N}) \cr
\end{bmatrix}.
\end{equation}
By  Farkas lemma, for example see \cite[Section 5.8]{Boyd},
the system (\ref{realsys})  has no non-trivial nonnegative solution
if and only if there is
a real vector $\bv = (a_0, \dots, a_N, b_0, \dots, b_N)$ such that
$\bv \tilde C_{\alpha,N}$ is a positive vector, i.e., all entries are
positive. Note that $a_0, a_1, \dots, a_N$
appear in $\bv \tilde C_{\alpha,N}$ as the $j$th entries for
$j = 1, 1+(N+1), 1+(N+1)+N, 1+(N+1)+N + (N-1),  \dots, (N+1)(N+2)/2$.
So, $a_0, \dots, a_N > 0$ if the said vector $\bv$ exists.
Set $\xi_j = a_j - ib_j$ for $j = 0, \dots, N$.
Then the system (\ref{realsys}) has no non-trivial nonnegative solution
if and only if condition (b) holds.
\qed

\section*{Acknowledgment}
The authors would like to thank
Runyao Duan, Cheng Guo, Yinan Li, and Weiyan Yu for some helpful correspondence and discussion.
Li is an affiliate member of the Institute for Quantum Computing,
University of Waterloo; he is also an honorary professor of Shanghai
University. His research was supported by USA NSF grant DMS 1331021,
Simons Foundation Grant 351047, and NNSF of China Grants 11571220 and 11971294.
The research of Liu was supported by the
National Natural Science Foundation of China grants: 11571075 and 11871015.
 The research of Ma was supported by the National Natural Science Foundation of China grants: 11601322 and 61573240.
 The research of Pelejo was supported by
UP Diliman OVCRD PhDIA 191902.

\nocite*

\end{document}